\documentclass[11pt,a4paper]{article}

\usepackage{a4wide}
\usepackage{graphicx}
\usepackage{hyperref}

\newcommand{\doi}[1]{\href{http://dx.doi.org/#1}{doi:#1}}
\newcommand{\identifiers}[2]{\href{http://identifiers.org/#1/#2}{#2}}
\newcommand{\tabref}[1]{Table~\ref{table:#1}}
\newcommand{\sbo}[1]{\href{http://www.ebi.ac.uk/sbo/main/SBO:0000#1}{#1}}

\begin{document}

\setlength{\parindent}{0pt}
\addtolength{\parskip}{6pt}
\newlength{\figWidth} \setlength{\figWidth}{0.5\textwidth}

\author{Kieran Smallbone \\ [24pt]
\emph{Manchester Centre for Integrative Systems Biology} \\
\emph{131 Princess Street, Manchester M1 7DN, UK} \\
\href{mailto:kieran.smallbone@manchester.ac.uk}{\tt kieran.smallbone@manchester.ac.uk}
}

\title{Standardized network reconstruction of CHO cell metabolism}

\date{}

\maketitle

\begin{abstract}

\noindent We have created a genome-scale network reconstruction of chinese hamster ovary (CHO) cell metabolism. Existing reconstructions were improved in terms of annotation standards, to facilitate their subsequent use in dynamic modelling. The resultant network is available from ChoNet~(\href{http://cho.sf.net/}{http://cho.sf.net/}).

\end{abstract}

\section*{ChoNet}

The structure of metabolic networks can be determined by a reconstruction approach, using data from genome annotation, metabolic databases and chemical databases~\cite{palsson10}. We built upon an existing reconstruction of the metabolic network of CHO cells that was based on genomic and literature data (Selvarasu \textit{et al.} \cite{selvarasu12}). This model contains 1065 genes, 1545 metabolic reactions, and 1218 unique metabolites. Use of \textit{in silico} modelling allows characterisation internal metabolic behaviour during growth and non-growth phases~\cite{selvarasu12}. 

Selvarasu \textit{et al.} suffers from the use of non-standard names and is not annotated with methods that are machine-readable. The model was thus updated according to existing community-driven annotation standards~\cite{herrgard08}. The reconstruction is described and made available in Systems Biology Markup Language (SBML) (\href{http://sbml.org/}{http://sbml.org/}, \cite{hucka03}), an established community XML format for the mark-up of biochemical models that is understood by a large number of software applications.  The network is available from ChoNet~(\href{http://cho.sf.net/}{http://cho.sf.net/}). As supplied, the model has an optimal growth rate of 0.0257 flux units.

\subsection*{Annotation}

The highly-annotated network is primarily assembled and provided as an SBML file. Specific model entities, such as species or reactions, are annotated using ontological terms. These annotations, encoded using the resource description framework (RDF)~\cite{wang05}, provide the facility to assign definitive terms to individual components, allowing  software to identify such components unambiguously and thus link model components to existing data resources~\cite{kell08}. Minimum Information Requested in the Annotation of Models (MIRIAM, \cite{lenovere05}) --compliant annotations have been used to identify components unambiguously by associating them with one or more terms from publicly available databases registered in MIRIAM resources~\cite{laibe08}. Thus this network is entirely traceable and is presented in a computational framework.

Six different databases are used to annotate entities in the network (see \tabref{1}). The Systems Biology Ontology (SBO)~\cite{courtot11} is also used to semantically discriminate between entity types. Five different SBO terms are used to annotate entities in the network (see \tabref{2}).

\begin{table}[!ht]
\begin{center}
\begin{tabular}{| c | c | c |}
	\hline 
	example 										& 	identifier				 			& 	database 			\\
	\hline 
	ChoNet										&	\identifiers{taxonomy}{10029}			&	taxonomy 			\\
	ChoNet										&	\identifiers{pubmed}{22252269}		&	pubmed 			\\
	cytosol 										&	\identifiers{obo.go}{GO:0005737}		&	obo.go			\\
	N-methylhistamine								&	\identifiers{chebi}{CHEBI:29009}		&	chebi			\\
	1-oxidoreductase								&	\identifiers{ec-code}{1.1.99.1}			&	ec-code			\\
	1-oxidoreductase								&	\identifiers{ncbigene}{218865}			&	ncbigene			\\	
	\hline
\end{tabular}
\caption{MIRIAM annotations used in the model.}
\label{table:1}
\end{center}
\end{table}

\begin{table}[!ht]
\begin{center}
\begin{tabular}{| c | c | c |}
	\hline 
	example 										& 	SBO term 		& 	interpretation			\\
	\hline 
	cytosol										&	\sbo{290} 		&	compartment			\\
	N-methylhistamine								&	\sbo{247} 		&	metabolite			\\
	N-methylhistamine								&	\sbo{176} 		&	biochemical reaction	\\
	AATRA20										&	\sbo{185} 		&	transport reaction		\\	
	biomass objective function						&	\sbo{397} 		&	modelling reaction		\\	
	\hline
\end{tabular}
\caption{SBO terms used in the model.}
\label{table:2}
\end{center}
\end{table}

\subsection*{Use}

We maintain the distinction between the CHO cell GEnome scale Network REconstruction (GENRE)~\cite{price04} and its derived GEnome scale Model (GEM)~\cite{feist08}. This is important to differentiate between the established biochemical knowledge included in a GENRE and the modelling assumptions required for analysis or simulation with a GEM. A GENRE serves as a structured knowledge base of established biochemical facts, while a GEM is a model which supplements the established biochemical information with additional (potentially hypothetical) information to enable computational simulation and analysis~\cite{heavner12}. Reactions added to the GEM include the biomass objective function -- a sink representing cellular growth -- and hypothetical transporters.

Three versions of the network are made available:

\begin{itemize}
\item \texttt{<organism>\_<version>.xml}, a GEM for use in flux analyses, provided in Flux Balance Constraints (FBC) format~\cite{fbc}
\item \texttt{<organism>\_<version>\_cobra.xml}, the same GEM network, provided in Cobra format~\cite{schellenberger11}
\item \texttt{<organism>\_<version>\_recon.xml}, a GENRE containing only reactions for which there is experimental evidence
\end{itemize}

\section*{EcoliNet and YeastNet}

EcoliNet~(\href{http://ecoli.sf.net/}{http://ecoli.sf.net/}) and YeastNet~(\href{http://yeast.sf.net/}{http://yeast.sf.net/}) are annotated metabolic network of Escherichia coli and Saccharomyces cerevisiae S288c, respectively, that are periodically updated by a team of collaborators from various research groups. The three networks are structured identically to facilitate comparative studies.

\paragraph{Acknowledgements}

This work is deliverable 4.3 of the EU FP7 (KBBE) grant 289434 ``\href{http://www.biopredyn.eu}{BioPreDyn}: New Bioinformatics Methods and Tools for Data-Driven Predictive Dynamic Modelling in Biotechnological Applications''.

\end{document}